\documentclass[a4paper,11pt]{article}

\pdfoutput=1

\usepackage[a4paper,left=2.73cm,right=2.7cm,top=3cm,bottom=3.5cm]{geometry}

\usepackage[T1]{fontenc} 
\usepackage{graphicx}
\usepackage{epsfig}
\usepackage{amssymb}
\usepackage{amsfonts}
\usepackage{dsfont}
\usepackage{amsmath,euscript,array,mathrsfs,nccmath}
\usepackage{multirow}
\usepackage{bbold,bm,bbm}
\usepackage{epsf}
\usepackage{slashed}
\usepackage{comment}
\usepackage{colortbl}
\usepackage{lmodern}
\usepackage[utf8]{inputenc}
\usepackage[noadjust]{cite}
\usepackage{changepage}

\usepackage{jheppub}

\usepackage{caption,subcaption,wrapfig}
\usepackage[colorlinks=true,linktocpage=true,linkcolor=blue,citecolor=blue]{hyperref}

\usepackage{tikz}
\usetikzlibrary{decorations.markings,decorations.pathmorphing}

\newcommand{\be}{\begin{equation}} \newcommand{\ee}{\end{equation}}
\newcommand{\bea}{\begin{eqnarray}} \newcommand{\eea}{\end{eqnarray}}

\newcommand{\dd}{\text{d}}

\newcommand{\BET}{\overline{b}}

\newcommand{\CP}{\mathbb{C}{\rm P}}

\newcommand{\cfh}{f_h}
\newcommand{\cgh}{g_h}
\newcommand{\clh}{\lambda_h}
\newcommand{\chh}{h_h}
\newcommand{\cbh}{\mathsf{b}_h}

\newcommand{\UU}{{\rm U}}
\newcommand{\SU}{{\rm SU}}
\newcommand{\USp}{{\rm USp}}

\newcommand{\newsec}[1]{\section{#1}}

\newcommand{\Bphys}{\mathsf{B}}
\newcommand{\Mphys}{\mathsf{M}}
\newcommand{\gs}{g_s}
\newcommand{\ls}{\ell_s}

\newcommand{\QD}{Q_{\text{\tiny D2}}}
\newcommand{\LQCD}{\Lambda_{\text{\tiny QCD}}}

\newcommand{\PreserveBackslash}[1]{\let\temp=\\#1\let\\=\temp}
\newcolumntype{C}[1]{>{\PreserveBackslash\centering}p{#1}}
\newcolumntype{R}[1]{>{\PreserveBackslash\raggedleft}p{#1}}
\newcolumntype{L}[1]{>{\PreserveBackslash\raggedright}p{#1}}

\begin{document}

	\begin{titlepage}
		
		\thispagestyle{empty}
		
		\begin{flushright}
			\hfill{NORDITA 2023-016}
		\end{flushright}
		
		\vspace{40pt}  
		
		\begin{center}

			\begin{adjustwidth}{-15mm}{-15mm}
				\begin{center}
					{\LARGE \textbf{ Baryonic matter at strong coupling:\\
							confining superfluids and deconfined ferromagnets}}
				\end{center}
			\end{adjustwidth} 
			
			\vspace{40pt}
			
			{\large \bf Ant\'on F. Faedo,$^{1,\,2}$   Carlos Hoyos,$^{1,\,2}$   \\ [1mm]
				
				and Javier G. Subils$^{3}$
			}

			\vspace{25pt}

			{\normalsize  $^{1}$ \textit{Departamento de F\'{i}sica, Universidad de Oviedo, \\ c/ Leopoldo Calvo Sotelo 18, ES-33007, Oviedo, Spain.}}\\
			\vspace{15pt}
			{ $^{2}$ \textit{Instituto Universitario de Ciencias y Tecnolog\'{\i}as Espaciales de Asturias (ICTEA), \\ Calle de la Independencia 13, ES-33004, Oviedo, Spain.}}\\
			\vspace{15pt}
			{ $^{3}$\textit{ Nordita, Stockholm University and KTH Royal Institute of Technology,\\
					Hannes Alfvéns väg 12, SE-106 91 Stockholm, Sweden.}}\\
			\vspace{15pt}

			\vspace{60pt}
			\textbf{Abstract}
		\end{center} 
		We study the phase diagram of a strongly coupled confining theory in $2+1$ dimensions, as a function of temperature and baryon chemical potential. The theory has a fully fledged supergravity holographic dual, that we use to predict a line of first order phase transitions separating a confining phase and a deconfined phase. Both phases exhibit a non-zero baryon density thus providing a first example of baryonic matter in a confining string dual that does not require the introduction of flavor branes. We argue that the confining phase is a baryon superfluid, while the deconfined phase has non-zero baryon magnetization.
	\end{titlepage}

	\newpage
	
	\tableofcontents

	\newsec{Introduction}
	\label{sec:intro}
	Quantum Chromodynamics (QCD) at a few times the nuclear saturation density, as found at the core of heavy neutron stars, is notoriously hard to describe. Perturbation theory is not applicable, lattice QCD suffers from the sign problem, and one is outside the regime of controlled effective theory descriptions such as chiral perturbation theory. Gauge/gravity duality avoids the aforementioned issues and may be useful to get a handle on this problem by studying theories similar to QCD at non-zero baryon density. However, it should be noted that an exact dual of QCD is not
	know, and tractable holographic models require additional
	simplifications of the field theory, such as taking parametrically
	large number of colors and 't~Hooft coupling, so one should be careful
	when extracting general lessons.
	
	A baryonic symmetry in holographic models is usually realized through flavor branes \cite{Karch:2003nh}.
	Baryon charge density is straightforward to introduce in deconfined phases, where the baryon charge in the dual description is inside the horizon of a black brane \cite{Kobayashi:2006sb,Mateos:2007vc,Bigazzi:2011it,Bigazzi:2013jqa,Bigazzi:2014qsa,Faedo:2015urf,Faedo:2017aoe}. In confined phases, however, it becomes technically challenging to deal with a finite density of baryons, since the dual objects are heavy solitons on the flavor branes \cite{Witten:1998xy,Brandhuber:1998xy,Sakai:2004cn,Hata:2007mb,Seki:2008mu}. This can be traced to the classical approximation in gravity corresponding to a large-$N$ limit in field theory, so that baryons are large operators with $\sim N$ fields. Nevertheless there have been attempts to construct finite baryon density states through soliton lattices \cite{Rho:2009ym,Kaplunovsky:2012gb,Bolognesi:2013jba,Kaplunovsky:2015zsa,Jarvinen:2020xjh} or by introducing additional phenomenological simplifications to deal with homogeneous configurations \cite{Bergman:2007wp,Rozali:2007rx,Kim:2007zm,deBoer:2012ij,Ghoroku:2012am,Li:2015uea,Preis:2016fsp,Elliot-Ripley:2016uwb,Elliot-Ripley:2016ctk,BitaghsirFadafan:2018uzs,Ishii:2019gta,Kovensky:2021ddl,Kovensky:2021kzl,Ghoroku:2021fos,Bartolini:2022rkl,CruzRojas:2023ugm} (see \cite{Hoyos:2021uff,Jarvinen:2021jbd} for recent reviews on the topic). Alternatively, one may look for different realizations that allow to introduce a non-zero baryon density in the gravity dual without the need for charged solitons. For instance, a possibility studied in  \cite{Hoyos:2016ahj} was to employ a Corrigan--Ramond-type of extrapolation to large-$N$ proposed in \cite{Hoyos-Badajoz:2009zmh}, but this was restricted to a quenched approximation for flavors. Here, we will follow a different path and present for the first time a fully fledged classical supergravity solution that is dual to a three-dimensional confining theory at non-zero baryon density in a homogeneous state and that does not rely on additional phenomenological assumptions. We will study the phase diagram as function of temperature and baryon chemical potential and identify the location of transitions to a deconfined phase. In contrast to other models, the flux dual to the baryon density is not sourced by introducing charged solitonic objects, but it is generated thanks to non-trivial interactions among neutral supergravity fields. A similar mechanism might allow to introduce baryon charge in supergravity duals to confining four-dimensional theories such as Klebanov--Strassler \cite{Klebanov:2000hb}. Black branes dual to deconfined phases with baryon charge were already found in \cite{Herzog:2009gd}.

	\newsec{Holographic model}
	\label{sec:setup}
	In the absence of baryon charge, the gravity dual geometry is the $\mathbb{B}_8^{\rm conf}$ solution identified in \cite{Cvetic:2001bw,Herzog:2002ss,Faedo:2017fbv} as dual to a three-dimensional confining theory that flows in the UV to a minimally supersymmetric Yang--Mills theory. At large enough temperatures there is a first order (Hawking--Page) transition to a black hole dual to a deconfined phase \cite{Elander:2020rgv}. These solutions can be found in a four-dimensional truncation of supergravity that allows for the inclusion of Abelian vector fields, one of which is massless and therefore dual to a global conserved current in the field theory. In \cite{Faedo:2022lxd} the family of solutions was extended by turning on the electric and magnetic components of the vector fields. Details about the solutions are given in Appendix A. With Dirichlet boundary conditions for the massless vector field, it was argued there that the dual was a quiver theory with $\UU(N)\times \UU(N+M)$ gauge group, with $N/M$ an integer, and the global current was identified as the topological current counting the magnetic flux of the diagonal $\UU(1)$ gauge group. The associated charge thus counts the number of monopoles. 
	
	The same class of gravity solutions has a different interpretation if an alternative set of boundary conditions is chosen for the vector field, following the discussion in \cite{Bergman:2020ifi}. With Neumann boundary conditions one can perform an electromagnetic duality transformation in the four-dimensional gravity theory, which from the three-dimensional field theory point of view is a particle-vortex or mirror duality transformation \cite{Witten:2003ya,Seiberg:2016gmd}. In this case the field theory becomes a quiver with $\SU(N)\times \SU(N+M)$ gauge group, and the global conserved current corresponds to a $\UU(1)_{\mathcal{B}}$ baryonic symmetry acting on fields in the bifundamental representation. In this map, global charge and background magnetic field are interchanged. Therefore, the confining solutions found in \cite{Faedo:2022lxd} with zero monopole charge and non-zero monopole magnetic field map through the particle-vortex duality to confining solutions with non-zero baryon charge and zero baryon magnetic field. 
	
	In order to complete the phase diagram we also construct black hole solutions dual to a deconfined phase with non-zero baryon chemical potential and zero baryon magnetic field. This can be done by starting with the black hole solutions found in \cite{Faedo:2022lxd} and tuning the monopole chemical potential (which was set to be zero in the original setup) until the monopole charge density vanishes at non-zero monopole magnetic field. We emphasize that all the solutions we consider are regular solutions to ten-dimensional type IIA supergravity.
	
	\subsection{Monopole-baryon duality}
	
	The ten-dimensional geometry dual to the three-dimensional confining theory is of the form ${\cal M}_4\times {\widetilde{\CP}}{}^3$, where ${\cal M}_4$ spans the field theory directions and the holographic radial coordinate, and ${\widetilde{\CP}}{}^3$ is an internal space which is a deformation of ${{\CP}}{}^3$. 
	
	Branes wrapped on the internal space and ending at the asymptotic boundary of ${\cal M}_4$ are dual to local operators. These branes couple electrically to a Ramond-Ramond form potential of rank equal to their dimension and magnetically to a form obtained through Hodge duality, see table below 
	
	\begin{center}
		\begin{tabular}
			{|C{3cm}|C{3cm}|C{3cm}|C{3cm}|}
			\hline
			RR-form & Hodge dual & electric & magnetic \\ \hline
			$C_1$ &  $C_7$ & D0 & D6 \\
			$C_3$ &  $C_5$ & D2 & D4 \\
			\hline
		\end{tabular}
	\end{center}
	
	D0 and D2 branes are dual to monopole operators, while D4 and D6 branes are dual to dibaryon and baryon operators respectively. Boundary conditions for the form potentials determine whether a brane can end at the boundary, and therefore, whether the corresponding dual local operator exists in the theory. A $\UU(N)\times \UU(N+M)$ theory has gauge-invariant monopole operators, but the would-be baryonic symmetry is gauged so there are no dibaryon or baryon operators in the physical spectrum. This translates into Dirichlet boundary conditions for $C_1$ and $C_3$. Conversely, a $\SU(N)\times \SU(N+M)$ theory contains baryon operators but no monopoles, and this translates into Dirichlet boundary conditions for $C_5$ and $C_7$. 
	
	Upon reduction to four-dimensional supergravity, the relevant components of each form introduce a vector field. More precisely, $C_1$ and $\int_{\widetilde{\CP}{}^3} C_7$ give rise to vectors $a_1$ and $A_1$ respectively. These are massless and thus holographically dual to conserved global currents. In particular, $a_1$ is dual to the monopole charge and $A_1$ to the baryon charge. Which symmetry is present is determined by the boundary conditions inherited from the forms in ten dimensions. We will identify the components with unit coupling to the D0 and D6 branes as the gauge fields dual to the monopole and baryon currents
	\begin{equation}
		a_1= g_s \ell_s \frac{M^2}{N} \mathsf{A}_{_{\cal M}},\ \  A_1=\frac{1}{3\cdot 2^5 \pi^3 g_s \ell_s^7} \frac{M}{N^2} \mathsf{A}_{_{\cal B}}\,,
	\end{equation}
	with $\gs$ and $\ls$ the string coupling and length.
	
	In \cite{Faedo:2022lxd} the reduction to four dimensions was made using $C_1$ and $C_3$, which introduces naturally Dirichlet boundary conditions for these forms, so the holographic dual contains monopoles. We can implement the change in boundary conditions that takes us to a theory with baryons by adding a term to the four-dimensional supergravity action
	\begin{equation}\label{eq:dualact}
		S_{\text{dual}}=\frac{1}{2\kappa_4^2}\int \, \dd a_1 \wedge \dd A_1=\frac{NM}{2\pi} \int \dd \mathsf{A}_{_{\cal M}} \wedge \dd \mathsf{A}_{_{\cal B}}\,, 
	\end{equation}
	where $\kappa_4^2$ is the four-dimensional Newton's constant. Since the new term is a total derivative, it does not affect the equations of motion or the solutions, but it modifies the value of the on-shell action.
	
	We introduce a similar term for the reduction of $C_3$ and $C_5$, so that D4 branes can end at the boundary and there are dibaryon operators in the holographic dual. This term, however, does not alter the value of the free energy because the corresponding gauge fields are massive, so we omit its explicit form here.
	
	\subsection{Baryon and dibaryon operators}
	
	The $\SU(N)\times \SU(N+M)$ theory is minimally supersymmetric in three dimensions. 
	It contains ${\cal N}=1$ scalar multiplets $A_1$ and $A_2$ in the bifundamental and $B^1$ and $B^2$ in the anti-bifundamental representations of the gauge group. They form doublets of the global flavor symmetry $\USp(2)\subset \SU(4)$, which matches the isometry of the internal space in the holographic dual. $A_i$ and $B^i$ also carry opposite charge under the global $\UU(1)_{\cal B}$ symmetry.
	
	Baryon and dibaryon operators are built in a way similar to the Klebanov--Strassler dual \cite{Gubser:1998fp,Aharony:2000pp}. A dibaryon operator is (omitting flavor indices)
	\begin{equation}
		D^{\beta_{N+1}\cdots \beta_{N+M}}=\epsilon^{\alpha_1 \cdots \alpha_{N+M}} \epsilon_{\beta_1\cdots \beta_N}A_{\alpha_1}^{\ \beta_1}\cdots  A_{\alpha_{N+M}}^{\ \beta_{N+M}}\,.
	\end{equation}
	The dibaryon is thus in a completely antisymmetric representation of the $\SU(N)$ group. The anti-dibaryon has a similar form with $B$s instead of $A$s. The baryon operator is a singlet of the gauge group. Its explicit form is
	\begin{equation}\label{eq:baryonop}
		{\cal B}=\epsilon_{\beta_1\cdots \beta_M \cdots \beta_{N-M+1} \cdots \beta_{N} }D^{\beta_1\cdots \beta_M}\cdots D^{\beta_{N-M+1}\cdots \beta_{N}}\,.
	\end{equation}
	Therefore, a baryon is a singlet formed out of $N/M$ dibaryons, and an anti-baryon is constructed in a similar way from anti-dibaryons.
	
	As we have mentioned, the dual of a dibaryon operator is a D4 brane wrapping a ${\CP}{}^2$ cycle in the internal space. The D4 Wess--Zumino action includes a term (see Appendix~B for details) 
	\begin{equation}
		S_{\text{D4}}\supset M\int {\cal A}\,, 
	\end{equation}
	where ${\cal A}$ is the gauge field on the brane and the $M$ factor is determined by the background $F_4$ flux. This produces a tadpole on the worldvolume that needs to be cancelled by attaching $M$ strings to the D4 brane, analogously to \cite{Aharony:2000pp}. The counterpart in the field theory is that a dibaryon operator is not gauge invariant but a gauge invariant operator can be formed by combining it with $M$ Wilson lines.
	
	The dual to the baryon operator, a D6 brane wrapping the whole $\widetilde{\CP}{}^3$, does not have tadpoles in agreement with ${\cal B}$ being a singlet. When the D6 brane is taken to the asymptotic boundary, the background $B_2$ field induces a coupling (see Appendix B) \footnote{This statement depends on the gauge choice for $B_2$. Through large gauge transformations this coupling can be set to zero, in which case the D6 develops a tadpole so that $N$ strings have to be attached to it. The baryon in this case is formed by adding $N/M$ D4 branes on which bundles of $M$ strings end.}
	\begin{equation}
		S_{\text D6}\supset  \frac{N}{M}\left(T_{\text D4}\int C_5\right)\,,
	\end{equation}
	with $T_{\text D4}$ the tension of a D4 brane. This implies that a D6 brane carries $N/M$ units of D4 brane charge, consistently with the expression in \eqref{eq:baryonop}. As the D6 is taken to the interior of the geometry, which corresponds to a flow from the UV to the IR of the field theory, there is a duality cascade that reduces the rank of the gauge groups, as described in \cite{Faedo:2022lxd}. The D4 charge carried by the D6 brane is also reduced through the cascade, until it becomes one when the brane is close to the origin, where the dual gauge group is just $\SU(M)$ and the theory confines. At this stage, the baryon and dibaryon operators are
	\begin{equation}
		D^{\beta_1\cdots \beta_M}=A^{[\beta_1}\cdots A^{\beta_M]},\ \ {\cal B}=\epsilon_{\beta_1\cdots \beta_M}D^{\beta_1\cdots \beta_M}\,.
	\end{equation}
	Therefore, in the IR we have just a ${\cal N}=1$ SQCD$_{2+1}$ theory with two flavors and the D6 brane is dual to the usual baryon.
	
	\newsec{Thermodynamics and phase diagram}
	\label{sec:thermo}
	
	\begin{figure}[t]
		\begin{center}
			\noindent\includegraphics[width=0.6\textwidth]
			{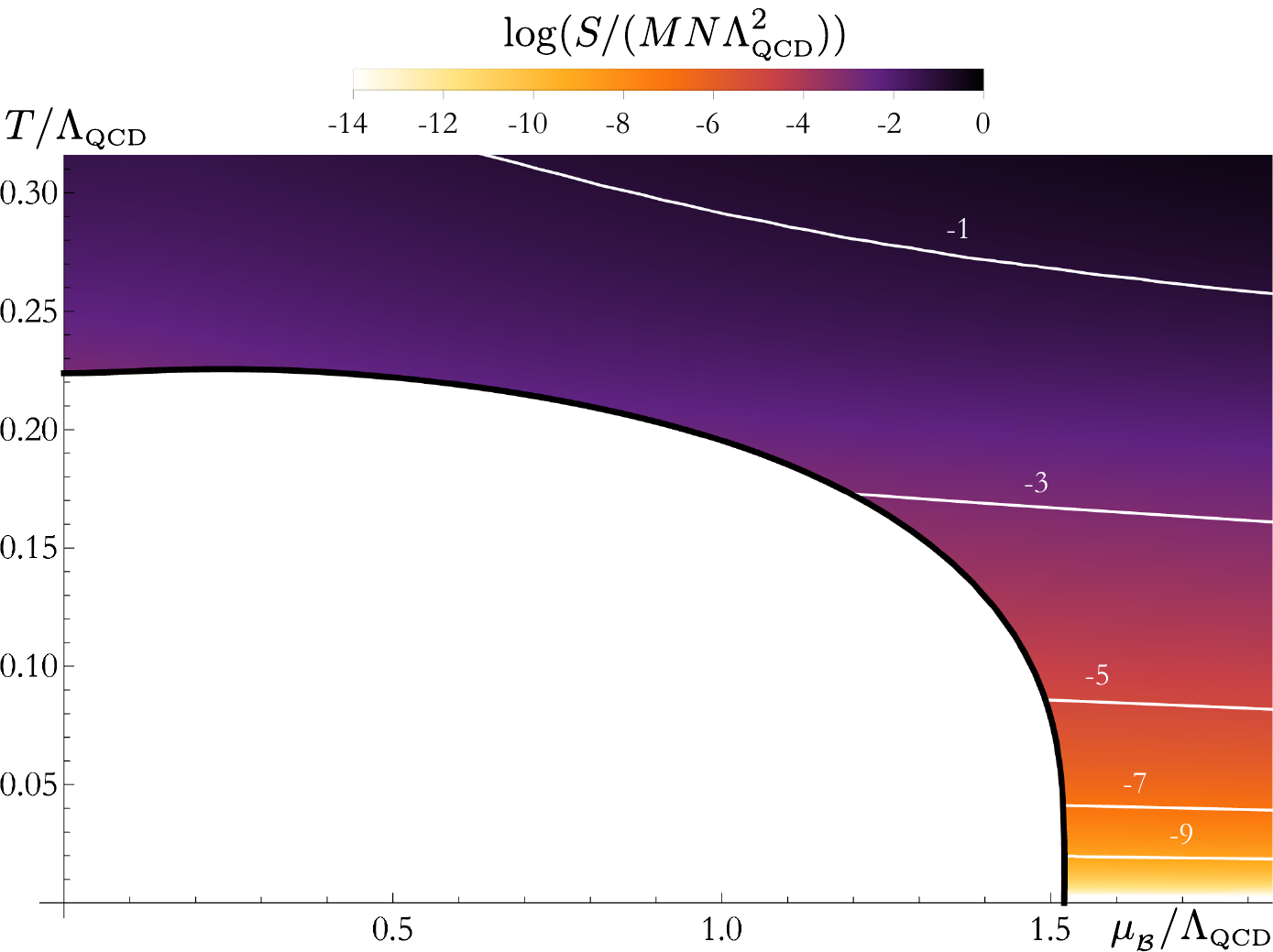}
		\end{center}
		\caption{\small  Density plot for the logarithm of the entropy density as a function of the chemical potential $\mu_{_{\cal B}}$ and temperature $T$. The confined and plasma phases are separated by a line of first-order phase transitions, represented by a thick black curve on the plot. Isentropic lines are shown in white.}
		\label{fig:entropy_density}
	\end{figure}
	
	Following the established holographic dictionary, the free energy equals the gravitational on-shell action. Expressions for thermodynamic quantities in the theory with monopoles, computed in \cite{Faedo:2022lxd}, are collected in Appendix C. They are given in units of the characteristic scale of confinement $\LQCD=\lambda (N/M)^3$, where $\lambda$ is the dimensionful 't Hooft coupling of $2+1$ super Yang--Mills.
	
\begin{figure}[t]
		\begin{center}
			\includegraphics[width=0.6\textwidth]
			{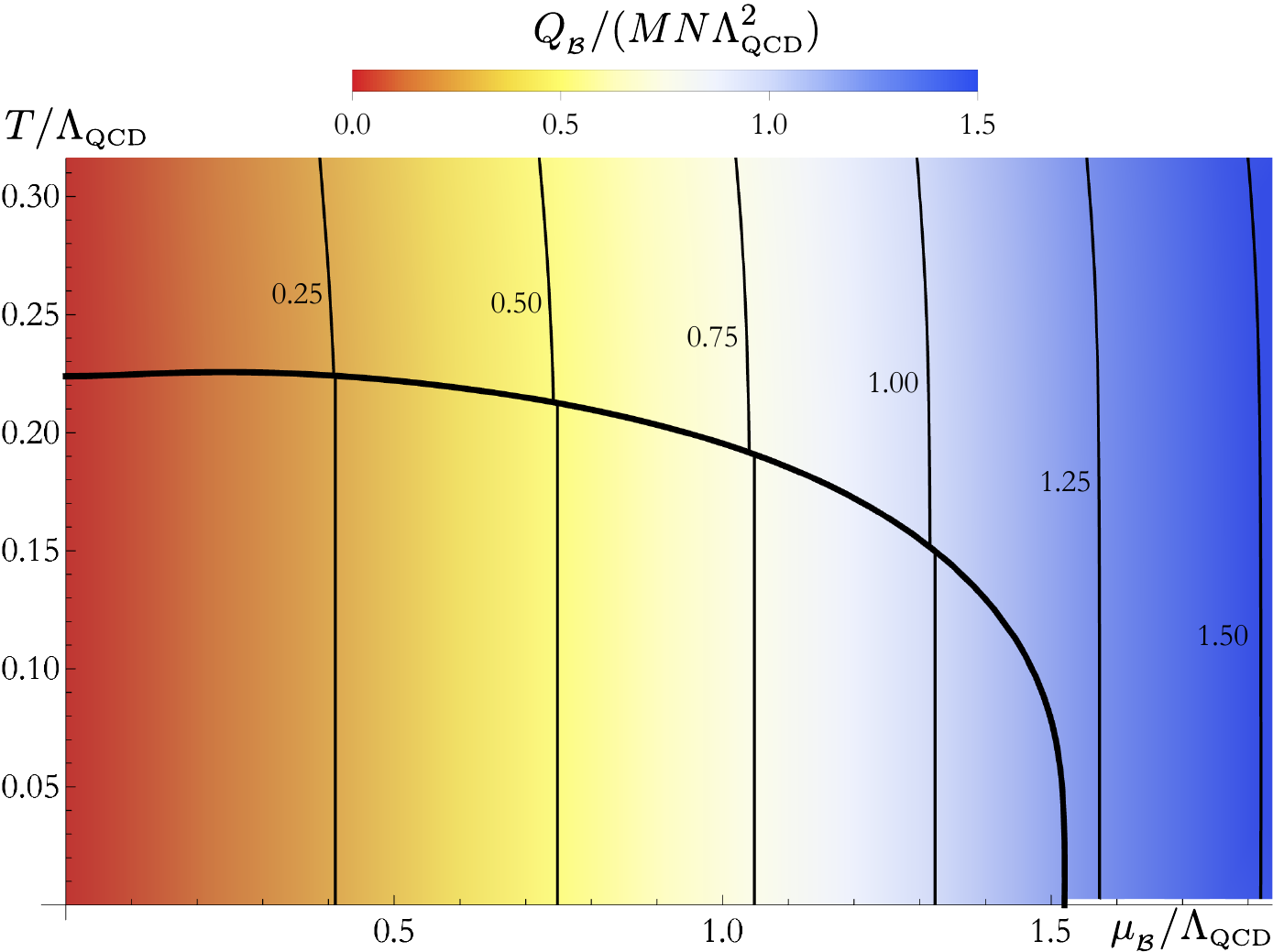}
		\end{center}
		\caption{\small  Density plot of the baryon charge of the preferred phase at every chemical potential and temperature. Remarkably, the charge density is finite in the confining phase. Contours of constant density are marked by thin black lines.}
		\label{fig:charge_density}
	\end{figure}

	The action of the theory with baryons differs only by the term in \eqref{eq:dualact}, which allows to derive a relation between the free energies of both theories. More precisely, if $\mu_{_{\cal M}}$, $\mu_{_{\cal B}}$ are the chemical potentials for monopoles and baryons and $Q_{_{\cal M}}$, $Q_{_{\cal B}}$ are the associated charges, the relation between the free energies is
	\begin{equation}
		G_{_{\cal B}}=G_{_{\cal M}}+\mu_{_{\cal M}}Q_{_{\cal M}}-\mu_{_{\cal B}} Q_{_{\cal B}}\,.
	\end{equation}
	The baryon chemical potential and charge are proportional to the monopole magnetization and magnetic field 
	\begin{equation}
		\mu_{_{\cal B}}=-\frac{2\pi}{NM} \frac{\partial G_{_{\cal M}}}{\partial \mathsf{B}_{_{\cal M}}},\qquad  \ Q_{_{\cal B}}=-\frac{NM}{2\pi}\mathsf{B}_{_{\cal M}}\,.
	\end{equation}
	There is a similar map between monopole charge and baryon magnetic field 
	\begin{equation}
		\mu_{_{\cal M}}= \frac{2\pi}{NM}\frac{\partial G_{_{\cal B}}}{\partial \mathsf{B}_{_{\cal B}}},\qquad \ Q_{_{\cal M}}=\frac{NM}{2\pi}\mathsf{B}_{_{\cal B}}\,.
	\end{equation}
	The equations and solutions for different values of the monopole charges and magnetic field were described in \cite{Faedo:2022lxd}. Here we can use those results together with the map between the theories with monopoles and baryons to describe the phase diagram of the latter. We will be focusing on the case with $\mathsf{B}_{_{\cal B}}=Q_{_{\cal M}}=0$. The results are summarized in figures~\ref{fig:entropy_density} and \ref{fig:charge_density}. We observe a line of first order phase transitions between a low temperature and chemical potential confined phase and a deconfined phase. Interestingly, the baryon density in the confined phase is non-zero. Thus, this corresponds to a genuine phase of homogeneous baryonic matter in holography. At the deconfining phase transition we observe that the entropy changes significantly, especially at higher temperatures, but the baryon density is quite insensitive. However, when the temperature goes to zero, the entropy density vanishes. The baryon density, though, still changes discontinuously and thus the transition remains of first order.
	
	In the following we will describe some of the properties of each phase.
	
	\subsection{Deconfined phase: baryon ferromagnet}
	
	In the deconfined phase there is a nonzero baryon magnetization at zero magnetic field, signalling parity breaking in the underlying theory. Since the baryon magnetic field has been set to zero, this implies that this phase is ferromagnetic. It should be noted that ``ferromagnetic'' in this context does not involve spin, or spontaneous breaking of a global symmetry. 
	
	The thermal equilibrium state can be thought of as a relativistic fluid at rest. When parity is broken there are additional transport coefficients contributing linearly in the magnetic field to the pressure \cite{Jensen:2011xb}. These can be extracted from a term of the (Euclidean) effective action that takes the simple form \cite{Banerjee:2012iz,Jensen:2012jh,Haehl:2013kra} 
	\begin{equation}
		G_{_{\cal B}}= \frac{T}{V_2}S_{\text{eff}}\supset-\frac{T}{V_2}\int \dd^3 x\, \mathsf{M}_{_{\cal B}}(T,\mu_{_{\cal B}})\, \mathsf{B}_{_{\cal B}}\,,
	\end{equation}
	where $V_2$ is the spatial volume. Charge conjugation symmetry forces the magnetization to be odd in the baryon chemical potential, $\mathsf{M}_{_{\cal B}}=-\mu_{_{\cal B}} \,\bar{\nu}(T,\mu_{_{\cal B}}^2)/(2\pi)$, in such a way that this term looks like an incomplete Chern--Simons term with coefficient $\bar{\nu}$.
	
\begin{figure}[t]
		\begin{center}
			\includegraphics[width=0.6\textwidth]
			{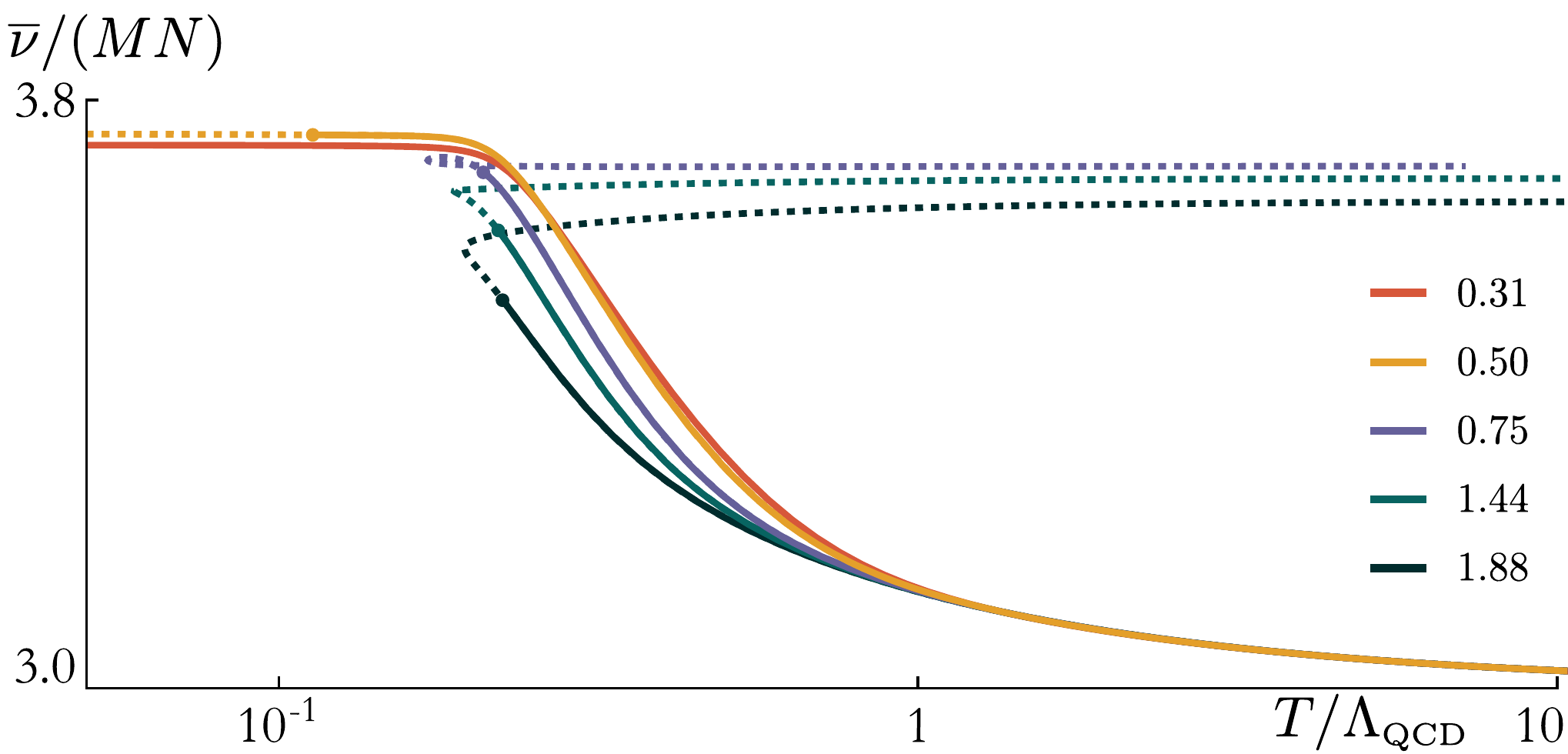}
		\end{center}
		\caption{\small  Coefficient $\bar\nu(T,\mu_{_{\cal B}}^2)$ appearing in the ratio between magnetization and baryon chemical potential, for different choices of the chemical potential, as a function of temperature in the deconfined phase. The solid (dashed) curves correspond to stable (unstable) phases. The dots mark the temperature of the phase transition. The corresponding values of $\mu_{_{\cal B}}/\LQCD$ are shown in the legend.}
		\label{fig:nuCoeff}
	\end{figure}

	We plot the value of the coefficient $\bar{\nu}$ as a function of the temperature for different chemical potentials in figure~\ref{fig:nuCoeff}. The coefficient shows a rather weak dependence on the baryon chemical potential, for the values we have studied, and it decreases from a finite value $\bar{\nu}\lesssim (3.6-3.8) \,NM$ as the temperature increases to an asymptotic value $\bar{\nu}(T\to \infty)=3\, NM$. This suggests that the UV theory has a parity anomaly that generates a Chern--Simons term for a background baryon gauge field. It should be mentioned that the supergravity flux dual to Chern--Simons terms for color gauge fields is zero in these solutions, so parity breaking should involve other fields, for instance through the generation of an effective fermion mass term.
	
	\subsection{Confined phase: baryon superfluid}
	
	In \cite{Faedo:2022lxd} it was shown that the monopole charge vanishes exactly in the confined phase. Following the particle-vortex duality, this implies that it is not possible to introduce a nonzero homogeneous magnetic field for baryons in the low temperature phase, a distinct signature of a superfluid baryon phase (which would be superconducting if the baryon gauge field was made dynamical). Since the superconductor is a perfect diamagnet, the magnetization will cancel exactly an applied magnetic field. Thus, the magnetization can take an arbitrary value as long as the superconducting phase is not destroyed. The equivalent statement in the dual, that the monopole chemical potential is arbitrary in the confined phase, was already pointed out in \cite{Faedo:2022lxd}.
	
	In this phase, the $\UU(1)_{\cal B}$ symmetry is spontaneously broken, so we expect to have a Goldstone boson, massless at zero $T$ and $\mu_{_{\cal B}}$. A similar situation was encountered in $3+1$ dimensions in the Klebanov--Strassler solution, where the supergravity mode dual to the Goldstone was identified in \cite{Aharony:2000pp,Gubser:2004qj,Gubser:2004tf}. In the case at hand we expect the Goldstone to be a mode of the
	vector fields.
	Note that Mermin--Wagner's theorem, which forbids symmetry breaking at finite temperature in $2+1$ dimensions, is avoided by the holographic model due to the large-$N$ limit that suppresses fluctuations.  
	
	\newsec{Conclusions}
	\label{sec:discuss}
	
	We have presented the first string theory example of a holographic dual to a strongly coupled confining theory with non-zero baryon density that does not rely on probe flavor branes and does not require considering multi-instanton solutions or phenomenological approximations to those. 
	
	In the realm of $2+1$ dimensions, we have applied particle-vortex duality to produce the phase diagram of a ${\cal N}=1$ supersymmetric theory with baryons, from the original one studied in \cite{Faedo:2022lxd}. These are close to non-supersymmetric theories in that there are no non-renormalization theorems based on holomorphycity, so that very little is known about them. For instance, the phase diagram has been studied previously only for theories with a single group factor or non-zero Chern--Simons levels \cite{Bashmakov:2018wts,Bashmakov:2021rci}.
	
	The theory exhibits rich dynamics, including a duality cascade and superfluid and (baryon) magnetic phases. It could be interesting to explore similar models for their possible application to condensed matter systems of strongly correlated fermions and unconventional superconductivity, see \cite{Blake:2022uyo} for a recent review on related topics.


	\vspace{0.3cm}
	\begin{acknowledgments}
		We thank David Mateos for useful discussions and comments on the manuscript.
		Nordita is supported in part by NordForsk. We thank the PDC Center for High Performance Computing, KTH Royal Institute of Technology,
		Sweden, for providing access to the computing resources used in this research (project
		SNIC 2021/22-999, 
		SNIC 2022/22-596). The work of A.F. and C.H is partially supported by the AEI and the MCIU through the Spanish grant PID2021-123021NB-I00 and by FICYT through the Asturian grant SV-PA-21-AYUD/2021/52177. A.F. is also supported by the “Beatriz Galindo” program, reference BEAGAL 18/00222.
	\end{acknowledgments}
	
	\appendix
	
	\setcounter{equation}{0}
	\renewcommand{\theequation}{A.\arabic{equation}}

	\section{Background}
	\label{app:metric}
	
	The metric describing the holographic dual geometry in type IIA supergravity is, in the string frame,
	\begin{eqnarray} \label{eq:ansatz_metric}
		\dd s_{\rm st}^2 &=&h^{-\frac12}\left(-\mathsf{b}\ \dd t^2 + \dd x_1^2 + \dd x_ 2^2\right) \notag\\ &+& h^{\frac12} \left(\frac{\dd r^2}{\mathsf{b}}+e^{2f}\dd\Omega_4^2+e^{2g}\left[\left(E^1\right)^2+\left(E^2\right)^2\right] \right)\,,\nonumber\\[2mm]
		e^\Phi&=&h^{\frac14} \, e^\Lambda \,,
	\end{eqnarray}
	with $f$, $g$, $h$, $\mathsf{b}$ and $\Lambda$ depending only on the radial coordinate $r$. Here, $E^1$ and $E^2$ are vielbein forms spanning a two-cycle in the internal space. Together with the $S^4$ metric $\dd\Omega_4^2$ they form a squashed $\CP^3$ (for $f=g$ one recovers the Fubini--Study metric and there is no squashing).
	
	There are two types of solutions: confining, where $e^{2g}\to 0$ at the origin (the two-cycle collapses smoothly) and black branes, where $\mathsf{b}\to 0$ at the horizon.
	
	In addition to the metric and dilaton, the following background forms are turned on
	\begin{equation}\label{eq:formansatz}
		\begin{aligned}
			F_4^{\rm fl} &=q_c\,\left(J_2\wedge J_2-X_2\wedge J_2\right)\,,\\[2mm]
			B_2 &= b_2+b_X\,X_2+b_J\, J_2\,,\\[2mm]
			C_1 &= a_1\,,\\[2mm]
			C_3 &= a_3+\tilde{a}_1\wedge X_2+\hat{a}_1\wedge J_2+a_J \, J_3\,,
		\end{aligned}
	\end{equation}
	where $J_2,X_2, X_3$ and $J_3$ are two and three-forms on the internal space, see \cite{Faedo:2022lxd} for more details. The functions $b_X$, $b_J$ and $a_J$ are scalars from the four-dimensional point of view and depend solely on the radial coordinate $r$ in \eqref{eq:ansatz_metric}. The flux $q_c$ determines the offset in the groups' ranks
	\begin{equation}\label{eq:qc}
		q_c = \frac{3\pi \ls^3 \gs}{4} \,M\,.
	\end{equation}

	There are three vectors (one-forms) that we parametrize as 
	\begin{equation}
		\begin{aligned}
			a_1 &= a_t\left(r\right)\,\dd t+\gs\ls\,\frac{M^2}{N}\,\frac{\Bphys_{_{\cal M}}}{2}\left(x_1\dd x_2-x_2\dd x_1\right)\,,\\
			\tilde{a}_1 &= \tilde{a}_t\left(r\right)\,\dd t\,,\qquad \hat{a}_1 = \hat{a}_t\left(r\right)\,\dd t\,,
		\end{aligned}
	\end{equation}
	for some constant $\Bphys_{_{\cal M}}$. The prefactor has been chosen so that it corresponds to the physical monopole magnetic field. We also have three- and two-forms $a_3$ and $b_2$. The only non-trivial components are
	\begin{equation}
		a_3=a_{t12}\left(r\right)\dd t\wedge\dd x_1\wedge\dd x_2\,,\ b_2=b_{12}\left(r\right)\dd x_1\wedge\dd x_2\,.
	\end{equation}
	Notice that the three-form lives in the external four-dimensional space and therefore it is non-dynamical. Indeed, it can be dualized to a constant $Q_c$ which regularity at the origin fixes to zero $Q_c=0$. Similarly $b_{12}$ can be dualized to an axion which gives mass to the combination of vectors $\tilde{a}_1-\hat{a}_1$, so it does not contain independent degrees of freedom.
	
	It is convenient to work with the radial coordinate
	\begin{equation}\label{eq:coordinate}
		\dd r = - \frac{\rho_0}{\xi^2\sqrt{1-\xi^4}}\dd\xi\,,
	\end{equation}
	where the scale is determined by
	\begin{equation}\label{eq:ratiobrho}
		\rho_0 =  |b_0| \frac{\ls^ 2}{2}\,\lambda\,\frac{M^2}{N^ 2}\,,
	\end{equation}
	with $b_0$ a dimensionless constant and $\lambda$ the dimensionful 't Hooft coupling of the dual theory. At vanishing magnetic field in the confining solutions its value is $b_0(\Bphys_{_{\cal M}}=0)=-3 K(-1)$, with $K(m)$ the complete elliptic integral of the first kind. This is also the value for black brane solutions. In this coordinate, the asymptotic boundary is at $\xi\to 0$ and the origin/horizon at $\xi\to 1$ for the confining solutions and at $\xi=\xi_h<1$ for the black brane solutions.

	\setcounter{equation}{0}
	\renewcommand{\theequation}{B.\arabic{equation}}
	
	\section{D4 and D6 actions}
	\label{app:baryons}
	
	The gauge theory is subject to a ``cascade'' that changes the rank of the groups as it flows towards the IR. The precise way in which the gravity dual captures this effect is explained in \cite{Faedo:2022lxd}. At each step of the cascade the NS two-form changes as
	\begin{equation}
		\delta B_2=\pi\ls^2\left(X_2-J_2\right)\,,
	\end{equation}
	with $\left(X_2-J_2\right)$ a closed but non-exact form - so that the equations of motion are not altered - that describes the two-cycle, with volume $\int_{\CP^1} (X_2-J_2)=4\pi$. This produces a change in the WZ action of a D6 brane wrapping the internal space 
	\begin{equation}
		\delta S_{\text{D6}} =- T_{\text D6} \int\delta B_2\wedge C_5 = - \left(T_{\text D4}\int C_5\right)\,,
	\end{equation}
	that is, at each step of the cascade the D6 brane looses one unit of D4 charge. There are $N/M$ steps in the cascade \cite{Faedo:2022lxd}, so one starts with $N/M$ units of dibaryon charge at the boundary and ends with vanishing charge at the bottom of the cascade. 
	
	On the other hand, the tadpole on the wrapped branes is measured by the corresponding Page charge. For a D4 brane it reads
	\begin{equation}
		T_{\text D4}\int F_4^{\rm fl}\wedge(2 \pi \ls^2\mathcal{A})=M\int\mathcal{A}\,,
	\end{equation}
	where we have used the four-form flux in \eqref{eq:formansatz} as well as the quantization condition \eqref{eq:qc}. This means that each D4 brane comes with $M$ strings attached in order to cancel the tadpole. 
	
	Similarly, the tadpole for a D6 brane vanishes in the IR but is $N/M$ at the boundary due to the cascade (see the computation of the Page charge in \cite{Faedo:2022lxd}). At every step of the cascade this tadpole is cancelled by the $M$ strings attached to each of the remaining D4 branes and ending on the D6.
	
	All of this is compatible with the interpretation of a D4 and a D6 brane as a dibaryon - which is not gauge invariant - and a baryon - a gauge-invariant combination of $N/M$ dibaryons - respectively.

	\setcounter{equation}{0}
	\renewcommand{\theequation}{C.\arabic{equation}}
	
	\section{Thermodynamic quantities}
	\label{app:thermo}
	
	The boundary expansions of the metric and dilaton take the form
	\be\label{eq:boundexpmet}
	\begin{split}
		&e^{2f}  \simeq  \frac{\rho_0^2}{2\xi^2} \left( 1 +\cdots  + f_5 \xi ^5   \right)\,,  \	e^{2g} \simeq \frac{\rho_0^2}{4\xi^2}\,,  \\	
		&\,\,\,\,\, h \simeq \frac{128\, q_c^2 }{15\rho_0^ 6}|b_0|\,\xi^5 \,,\ \mathsf{b} \simeq  1  + \mathsf{b}_5 \xi^5 \,, \ e^\Lambda\simeq  1\,,
	\end{split}
	\ee
	where we have showed only the leading terms and the independent subleading coefficients appearing later in the expressions for thermodynamic quantities. Similarly, for the scalars the expansions are 
	\be\label{eq:boundexpscal}
	b_J \simeq \frac{2q_c}{3\rho_0 }\, b_0\,, \  b_X \simeq -  \frac{2q_c}{3\rho_0 }\, b_0 \,, \ a_J \simeq \frac{q_c}{6} \,.
	\ee
	Finally, the vector potentials are written as
	\begin{equation}\label{eq:boundexpgauge}
		a_t \simeq \frac{\rho_0^3}{q_c}  \, (v_0+v_1\xi),  \	\hat{a}_t \simeq \rho_0^2 \frac{2b_0v_1}{15}\xi , \	\tilde{a}_t \simeq -\rho_0^2 \frac{2b_0v_1}{15}\xi  \,. 
	\end{equation}
	The boundary expansion of the warp factor determines the number $N$ of D2-branes as follows
	\begin{equation}
		h \simeq \frac{16}{5}\, \frac{\QD}{r^ 5}\,,\qquad \qquad \QD = 3\pi^2 \ls^5 g_s N\,.
	\end{equation}
	With $N$ determining the rank of the gauge groups.
	
	We will also use the expansions at the horizon in the black brane solutions. The leading terms are
	\begin{equation}\label{eq:exphormet}
		\begin{split}
			&e^{2f} \simeq \rho_0^2\,  \cfh\,,  \ e^{2g} \simeq \rho_0^2 \,\cgh\,,  \
			h \simeq \frac{128\, q_c^2 }{9\rho_0^ 6}\, \chh \,,\\ &\mathsf{b}\simeq \cbh (\xi-\xi_h)\,,\  e^{\Lambda}\simeq \clh\,.
		\end{split}
	\end{equation}
	
	It will be convenient to introduce the characteristic scale of the confining phase
	\be\label{eq:lambdaQCD}
	\LQCD=\lambda\left(\frac{M}{N}\right)^3\,.
	\ee
	The map to thermodynamic quantities in the theory with monopoles was derived in \cite{Faedo:2022lxd}, here we present a summary of the results.
	In the black brane solutions (plasma phase) the temperature and entropy density are obtained from the surface gravity and area of the horizon
	\begin{equation}
		\begin{aligned}\label{eq:entropy_temperature}
			s_{\text{\tiny plas}} &  =NM\LQCD ^ 2\frac{|b_0|^3}{2^{11}\pi^3} \, \frac{512 \sqrt{2} \pi  \cfh^2 \cgh \chh^{\frac{1}{2}}}{3 \clh^2}\,,\\[2mm]
			T_{\text{\tiny plas}}&=  \LQCD\frac{b_0^2}{3\pi } \,  \frac{3 |\cbh| \xi_h^2 (1-\xi_h^4)^{\frac{1}{2}}}{32 \sqrt{2} \pi \chh^{\frac{1}{2}}}\,. 
		\end{aligned}
	\end{equation}
	The entropy vanishes in the confining phase $s_{\text{\tiny conf}} = 0$, while the temperature $T_{\text{\tiny conf}}$ is arbitrary.
	
	The chemical potential is fixed by the asymptotic value of $\mathsf{A}_{_{\cal M}\,t}$. For the black brane solutions it is 
	\begin{equation} \label{eq:chemical potential}
		\mu_{_{\cal M}} =\LQCD\frac{|b_0|^3}{6\pi}\,v_0\,.
	\end{equation}
	In the confining solutions $\mu_{_{\cal M}}$ is arbitrary. The monopole charge density is
	\begin{equation}\label{eq:charge}
		Q_{_{\cal M}} =-NM\LQCD ^ 2\frac{b_0^4}{2160\pi^3}(20 \BET b_0 + 27 v_1)\,.
	\end{equation}
	The monopole charge density vanishes identically when evaluated on confining solutions, but it is generically non-zero in the plasma phase.
	
	The energy density, pressure and spatial components of the energy momentum tensor are
	\begin{equation} \label{eq.holoTmunu}
		\begin{aligned}
			{E}_{_{\cal M}}  & =   NM\LQCD ^ 3 \, \frac{|b_0|^ 5}{3\cdot 2^{11} \pi^ 4}\left(-\frac{7 \mathsf{b}_5}{2}\, -  f_5 \right)  \,, \\[2mm] 
			T^x_{\ x}   &=  {P}_{_{\cal M}} - \Bphys_{_{\cal M}}\Mphys_{_{\cal M}} \\ & = NM\LQCD ^ 3 \, \frac{|b_0|^ 5}{3\cdot 2^{11} \pi^ 4} \left( -\frac{3 \mathsf{b}_5}{2}\, +  f_5\right)\,, \\[2mm]
		\end{aligned}
	\end{equation}
	were $\Mphys_{_{\cal M}}$ is the monopole magnetization. 
	
	Finally, the magnetization can be written as
	\begin{equation}\label{eq.magnetization}
		\begin{aligned}
			\Mphys_{_{\cal M}\,\text{\tiny conf}}  &= NM\LQCD \, \frac{3 \cdot  5\, }{2^ {10} \pi^2 }   \frac{\mathsf{b}_5}{\BET}\,,\\[2mm]
			\Mphys_{_{\cal M}\,\text{\tiny plas}}  &=NM\LQCD \, \frac{3 \cdot  5\, }{2^ {10} \pi^2 }   \left[\frac{\mathsf{b}_5}{\BET} -\frac{256}{135} b_0^3 v_0\right.\\ & \left.- \frac{16}{25} \frac{ 5 \cbh \cfh^2 \cgh \xi_h^2 (1-\xi_h^4)^{\frac{1}{2}}+4 b_0^2 \clh^2v_0 v_1}{\BET \clh^2}\right]\,,
		\end{aligned}
	\end{equation}
	for the confined and plasma phases respectively.

	\bibliographystyle{apsrev4-2}
	\bibliography{bibfile.bib}
	
\end{document}